\documentclass[useAMS,usenatbib]{mn2e}
\usepackage{epsfig}
\usepackage{amsmath}
\usepackage{amssymb}
\title[LISA detection of SMBH coalescence]{Realistic Event Rates for
  Detection of Supermassive Black Hole Coalescence by LISA}
\author[K.J. Rhook \& J.S.B. Wyithe]{Kirsty J.
  Rhook$^{1}$\thanks{krhook@ast.cam.ac.uk} \& J. Stuart B.
  Wyithe$^{2}$\thanks{swyithe@isis.ph.unimelb.edu.au}\\$^{1}$Institute
  of Astronomy, Madingley Road, Cambridge, CB3 OHA\\$^{2}$University of
  Melbourne, Parkville, Victoria, Australia}

\begin{document}

\date{Recieved 2004 April 26; Accepted 2005 March 3}

\pagerange{\pageref{firstpage}--\pageref{lastpage}} \pubyear{2005}
  
\maketitle

\label{firstpage}

\begin{abstract}
The gravitational waves generated during supermassive black hole
(SMBH) coalescence are prime candidates for detection by the satellite
LISA.  We use the extended Press-Schechter formalism combined with
empirically motivated estimates for the SMBH--dark matter halo
mass relation and SMBH occupation fraction to estimate the maximum
coalescence rate for major SMBH mergers.  Assuming efficient binary
coalescence, and guided by the lowest nuclear black hole mass inferred
in local galactic bulges and nearby low-luminosity active galactic
nuclei ($\approx 10^5$~M$_\odot$) we predict approximately 15
detections per year at a signal to noise greater than five, in each of
the inspiral and ringdown phases. Rare coalescences between SMBHs
having masses in excess of $10^7$~M$_\odot$ will be more readily
detected via gravitational waves from the ringdown phase.
\end{abstract}

 
\section{Introduction}

There is evidence for the existence of supermassive black holes
(SMBHs) over a range of epochs; SMBHs with masses $10^ 5 -
10^9$~M$_\odot$ are ubiquitous in local galactic bulges
(e.g. Magorrian et al.~1998: Ferrarese~2002), while the SMBHs powering
quasars [which have been detected at redshifts as high as $z\sim6 $
(Fan et al.~2001; Fan et al.~2003)] are estimated to range between
$10^6 - 10^9$ M$_\odot$ (see e.g. Shields et al. 2003). Evidence for
the early assembly of SMBHs, when interpreted within a hierarchical
cosmology, suggests that SMBH coalescence may be a frequent event.  If
this is the case, the gravitational waves (GWs) generated during SMBH
coalescence are a prime candidate for detection by the \emph{Laser
Interferometer Space Antennae}
(LISA\footnote{http://lisa.jpl.nasa.gov/}, Folkner et al.~1998;
Flanagan \& Hughes~1998; Hughes et al.~2001).

Estimates of the SMBH coalescence rate depend crucially on the
occupation fraction of SMBHs in halos, and therefore on the adopted
model for the formation and growth of SMBHs.  Within a hierarchical
CDM cosmology, both seeded SMBH formation models [in which a
primordial population of SMBHs grow via accretion and/or merging
(eg. Volonteri et al. 2003)] and ongoing formation models [where SMBHs
form jointly with bulges in halos (eg. Kauffmann \& Haehnelt 2000)]
are consistent with the present-day ubiquity of SMBHs in galactic
bulges (Menou et al. 2001).  

The formation of a SMBH may be limited to potential wells above a
minimum depth. If so, then there exists a critical halo virial
temperature ($T_{\rm vir,min}$) below which a halo cannot host SMBH
formation. If $T_{\rm vir,min}$ is low (eg. $T_{\rm vir,min} \approx
10^5$~K), then SMBHs are abundant in small halos, and predicted event
rates are in the order of 100's per year (Wyithe \&
Loeb 2003; Haehnelt 2004). This event rate may be significantly lower, less than one
per year, if $T_{\rm vir,min} \approx 10^6$~K and SMBHs form only at
the centres of massive galaxies (Haehnelt 2004). Hence the detection
rate has the potential to constrain the global SMBH population [with
the important caveat of the unknown efficiency of binary black hole
(BBH) coalescence, see \textsection \ref{SNR}].

The rate of coalescence detectable by LISA depends on the form of the
GW signal and the instrument's sensitivity curve. Previous estimates
of the event rate have used characteristic (rather than
time-dependent) GW signals to determine approximate detection limits
(Wyithe \& Loeb~2003) or have estimated the number of detectable
events by comparing the sensitivity curve with the total gravitational
wave background due to coalescing BBHs (Sesana et al.~2004a).
Recently Sesana et al.~(2004b) estimated the expected event rate for
a detailed physical model for SMBH growth. We perform the first
empirically motivated calculations for the LISA detection rate of SMBH
mergers in a hierarchical cosmology that use accurate signal to noise
ratios (SNRs) to determine detection criteria. In \textsection
\ref{SNR} we describe the calculation of SNRs for the LISA detection
of BBH coalescence.  In \textsection \ref{merger} we discuss our
calculation of the SMBH merger rate, including our halo merger rate
predictions, estimate for the occupation fraction of SMBHs in halos
and the $M_{\rm bh}$--$M_{\rm halo}$ relationship. Finally, we present
our event rate predictions as a function of $T_{\rm vir,min}$
(\textsection \ref{merger}) before concluding in
\textsection\ref{concl}.  Throughout this work we assume
$\Omega_m=0.27$, $\Omega_k=0$, $\Omega_{\Lambda}=0.73$,
$H_0\equiv100h$~km~s$^{-1}$~Mpc$^{-1}=71$~km~s$^{-1}$~Mpc$^{-1}$,
$\sigma_8=0.84$ and a primordial power spectrum with slope $n=1$ as
determined by the \emph{Wilkinson Microwave Anisotropy Probe} (WMAP,
Spergel et al. 2003).

\section{Binary black hole coalescence and detection by LISA}\label{SNR}

The efficiency with which BBHs coalesce is highly
uncertain. Initially, the SMBHs sink independently toward the centre
of a merged system due to dynamical friction from the dark matter
background until they form a bound binary (Begelman, Blandford \&
Rees 1980). The efficiency of this process depends on the orbital
parameters of the merging dark matter halos (van den Bosch et
al. 1999; Colpi, Mayer \& Governato 1999).  As the orbital separation
($r$) decreases, 3-body interactions with stars that pass within
$\approx r$ of the BBH centre of mass (the ``loss cone'') increasingly
dominate the energy loss.  Depending on the orbital parameters of the
binary and the background distribution of stars, this process may
result in a hardened [orbital velocity ($v$) $\sim$ stellar velocity
dispersion ($\sigma$)] binary system.  If hardening continues until a
binary separation where energy losses are dominated by GWs, the binary
will coalesce.

Yu (2002) argues that the efficiency of BBH coalescence hinges on the
time-scale for BBH hardening during the hard binary phase.  During
this stage, three-body interactions between the BBH and individual
stars eject stars into highly elliptical orbits, lowering the inner
stellar density and slowing further hardening (eg. Volonteri et
al. 2003).  The deceleration of the hardening rate is compounded by
the preferential depletion of stars within the loss cone (region of
parameter space where angular momentum is low enough that stars pass
near the BBH), resulting in a deficiency of stars that can extract
energy from the binary system (see Yu~2002).
 
Uncertainties in the efficiency of processes which may replenish loss
cone stars make this effect difficult to analyse.  Two-body stellar
relaxation is expected to result in some diffusion of stars back into
the loss cone (Binney \& Tremaine 1987) but it is unclear that this
process alone can support sufficient hardening. Numerical N-body
simulations have been used to include the effect of the BBH 'wander'
within the star field (due to 3-body interactions with stars), which
increases the effective size of the loss cone and may prevent the
coalescence from stalling (Quinlan \& Hernquist 1997; Milosavljevi{\'
c} \& Merritt 2001; Chatterjee, Hernquist \& Loeb 2003).  Other
scenarios which may aid evolution of the BBH into the GW dominated
regime have been proposed, including the effects of gaseous disks
(Gould \& Rix 2000), flattened or triaxial stellar distributions (Yu
2002) and disruption by a third SMBH (Hut \& Rees~1992).

The efficiency of BBH coalescence is of prime concern for LISA
detection rates.  Whilst undoubtably still an open question, we note
that there are plausible mechanisms to extract the required energy
  from the binary system.  For the purposes of our calculation
(\textsection\ref{merger}) we therefore assume that hardening of the
binary to the GW dominated regime occurs with an efficiency
($\epsilon_{\rm mrg}$) of unity. All event rates quoted in this paper
are proportional to $\epsilon_{\rm mrg}$.

The gravitational wave dominated portion of BBH coalescence may be
divided into three main phases (Hughes~2002). The binary begins in the
\emph{inspiral phase}, with the SMBHs slowly spiralling into tighter
orbits due to the adiabatic loss of GW energy. Eventually the dynamics
become relativistically unstable and the SMBHs violently plunge to
form a single object (\emph{merger phase}). The final GW signal can be
described by modelling the merged system as a perturbed Kerr SMBH
(\emph{ringdown phase}). The dynamics of the inspiral and ringdown
phases are well understood and theoretical waveforms for the 
GWs have been derived. However,
 certain parameters of the ringdown solution depend on unknown details
of the merger phase and must be guided by the results of numerical
simulation (see \textsection \ref{ringdown}).

Detection of GWs from a single binary source will be complicated by
the presence of galactic and extra-galactic GW foregrounds, and by
LISA's sensitivity to GWs from all sky directions. Matched filter
template searches will therefore be necessary to detect the GW signal
from an individual event (Hughes et al. 2001). The SNR for a matched
filter detection ($\rho$) is defined by the ratio of the coherently
folded signal and noise powers:

\begin{equation}
{\rho}^2 = 4\int_{0}^{\infty} \frac{|\tilde{h}(f)|^2}{S_h(f)}df,
\label{sndefn}\end{equation}

\noindent where $\tilde{h}(f)= \int_{-\infty}^{\infty} e^{2 \pi i f t}
h(t)dt$ is the Fourier transform of the dimensionless strain and
$S_h(f)$ is the spectral power of the noise. We calculate SNRs for the
final year (before the merger phase) of the inspiral signal.  The
effective duration of the ringdown phase is much shorter than one year
and we therefore calculate ringdown SNRs over the entire phase.
Estimates of LISA's noise spectrum are evolving with plans for the
instrument design.  A recently suggested sensitivity goal
(Bender~2003) has a (dimensionless) threshold sensitivity [defined as
$\frac{5S_h(f)^{1/2}}{\sqrt{5\times3.15\times10^{7}{\rm s}}}$, for
$\rho = 5$ and 1 yr of observation] with a low frequency power-law
slope of $-2.5$ for frequencies ranging between $10^{-5}$ and $
10^{-4}$~Hz, and a power-law slope of $-3$ for frequencies of $3
\times 10^{-6}$ to $10^{-5}$~Hz. Bender~(2003) suggests a hard low
frequency cut off at $f_{\rm min} = 3 \times 10^{-6}$~Hz, roughly
corresponding to the lowest resolvable frequency for a year long
mission. LISA will be sensitive to frequencies up to $f_{\rm max}
\approx
1$~Hz\footnote{http://www.srl.caltech.edu/\%7Eshane/sensitivity/}. As
a general rule $\rho \ge 5$ is required for the confident detection of
a signal (Hughes et al.~2001).

\subsection{Inspiral phase}

As the orbital separation of the binary shrinks, both the amplitude
and frequency of the inspiralling strain increase. To calculate
accurate inspiral phase SNRs, it is important to include the
frequency dependence of the strain in the integration of signal to
noise.  We achieve this using the technique outlined by Flanagan \&
Hughes (1998, henceforth FH98). FH98 show that sky averaged squared SNR
($\langle \rho^2 \rangle)$ can be re-expressed in terms of the emitted
GW energy spectrum $\frac{dE_e}{df_e}$

\begin{equation}\label{en_spec_snr}
\langle \rho^2 \rangle =\frac{2}{5 \pi^2
D_c(z)^2}\frac{G}{c^3}\int_{0}^{\infty}\frac{1}{f^2
S_h(f)}\frac{dE_e}{df_e}(f)df,
\end{equation}

\noindent where $f_e$ is the frequency of the GW. This is related to
the observed GW frequency ($f$) via the cosmological redshift of the
source [$f = f_e/(1+z)$]. The inspiral energy spectrum may be easily
calculated to first order from the quadrupole approximation for the
gravitational wave luminosity (FH98)

\begin{eqnarray}\label{insp_spec}
\nonumber
\frac{dE_{\rm e}}{df_{\rm e}}&=&\frac{1}{3}{\pi}^{2/3}G^{2/3}
 \frac{M_{\rm bh} \Delta M_{\rm bh}}{(M_{\rm bh} +\Delta M_{\rm
 bh})^{1/3}}f_e^{-1/3},\\ \hspace{-10mm}&=& \frac{1}{3}{\pi}^{2/3}G^{2/3} \mu
 M_{\rm total}^{2/3}f_e^{-1/3},
\end{eqnarray}

\noindent where $M_{\rm bh}$ and $\Delta M_{\rm bh}$ are the
individual BBH masses corresponding to a reduced mass $\mu =
\frac{M_{\rm bh} \Delta M_{\rm bh}}{M_{\rm bh} + \Delta M_{\rm bh}}$
and total mass $M_{\rm total}= M_{\rm bh}+\Delta M_{\rm bh}$.
Numerical simulations of BBH coalescence indicate that this
approximation is reasonable until $f$ reaches $f_{\rm merge}(z) =
205(1+z)^{-1}(M_{\rm total}/20{\rm M}_\odot)^{-1}$~Hz (FH98) which
roughly corresponds to the observed frequency when equal mass BBHs
have an orbital separation of 3 Schwarzchild radii ($\frac{6G M_{\rm
bh}}{c^2}$). This frequency forms the upper limit of integration in
equation (\ref{en_spec_snr}). Since the GW frequency of BBH
coalescence typically increases with time, this frequency may be used
to delineate the inspiral and merger phases.  The observed GW
frequency of the inspiral signal at an observed time $t^{\prime}$
prior to the merger phase is given by (FH98)

\begin{eqnarray}\label{f_oneyr}
\nonumber
f(t^{\prime})^{-8/3} &=& f_{\rm merge}(z)^{-8/3}\\&+& \frac{64
           \pi^{8/3}}{5} \frac{G^{5/3}}{c^5} [M_{\rm
           total}(1+z)]^{5/3} t^{\prime}.
\end{eqnarray}

\noindent Depending on the binary characteristics, $f_{\rm oneyr}
\equiv f(t^{\prime} = 3.15 \times 10^7{\rm s})$ may lie below the LISA
waveband. We therefore use the greater of $f_{\rm oneyr}$ and $f_{\rm
  min}$ as the lower limit of integration in
equation~(\ref{en_spec_snr}).

\subsection{Ringdown phase}\label{ringdown}

We use the technique described by Hughes (2002) to calculate SNRs for
the ringdown phase. BBH coalescence is expected to result in a
rotating SMBH with a bar-like excitation of the event horizon
(Hughes~2002).  This distortion of the Kerr solution may be modelled
by the bar-like [$(l,m)=(2,2)$] quasi-normal mode (Kokkotas \& Schmidt
1999).  The corresponding time dependent strain is described by an
exponentially damped sinusoid with total (observed) amplitude
($\mathcal{A}_{\rm rd}$) whose value depends on the fraction of the
BBH mass-energy ($M_{\rm total}c^2$) radiated during the ringdown
phase ($\epsilon_{\rm rd}$), the binary mass, and the spin parameter of
the merger product ($a$). The amplitude is
\begin{equation}\label{amp}
\mathcal{A}_{\rm rd}=\left( \frac{G}{c}
\right)^{1/2}\frac{1}{D_c(z)}\sqrt{\frac {5 \epsilon_{\rm rd}M_{\rm
total}}{4 \pi f_{\rm rd}Q}} \left[ \frac{4\mu}{M_{\rm total}} \right]
^2,
\end{equation}
\noindent  where $D_c(z)$  is the  comoving  distance to  a source  at
redshift $z$. The quality factor  ($Q$) is related to the damp-time of
the GW signal ($\tau$) by $Q=  \pi \tau f_{\rm rd}$.  The final factor
in equation~(\ref{amp}) accounts for  the reduced signal amplitude for
unequal  mass BBH  coalescence (FH98).  Numerical  simulations suggest
that $\epsilon_{\rm  rd} \approx  0.01 - 0.03$  (Baker 2001;  FH98 and
references therein).   We adopt  $\epsilon_{\rm rd} =  0.01$. Assuming
that   the   distribution   of   the  strain   between   polarisations
($h_+,h_{\times}$) mimics  that of the inspiral  phase, the dependence
of $h_+$ and  $h_{\times}$ on the orientation of  the system's angular
momentum  ($\hat{L}$)  and  sky position  [$\hat{n}(\theta,\phi)$]  is
given by (Fryer, Holtz \& Hughes 2002)
\begin{eqnarray}
\nonumber h_+(t)&=&\mathcal{A}_+{\rm exp}(-\pi f_{\rm rd,obs}t/Q){\rm cos}(2 \pi f_{\rm rd,obs}t + \lambda);\\ \nonumber
h_{\times}(t)&=&\mathcal{A}_{\times}{\rm exp}(-\pi f_{\rm
rd,obs}t/Q){\rm sin}(2 \pi f_{\rm rd,obs}t + \lambda),
\end{eqnarray}
where
\begin{eqnarray}
\nonumber \mathcal{A}_+&=&\mathcal{A}_{\rm rd}[1+{(\hat{L}.\hat{n})}^2];\\ \nonumber
\mathcal{A}_{\times}&=&-2 \mathcal{A}_{\rm rd}(\hat{L}.\hat{n}),
\end{eqnarray}

\noindent $f_{\rm rd,obs} = \frac{f{\rm rd}}{1+z}$ is the observed
ringdown GW frequency and $\lambda$ is the initial phase.  Leaver
(1985), Echeverria (1989) and Fryer Holtz \& Hughes (2002) have
produced fitting formulae for $f_{\rm rd}$ and $Q$ in terms of $M_{\rm
total}$ and $a$:
\begin{equation}\label{rd_fits1}
f_{\rm rd}=\frac{10^{5.3}}{2\pi} {\rm Hz}\left( \frac{M_{\rm total}}{{\rm
M}_{\sun}}\right)^{-1}[1-0.63(1-a)^{3/10}],
\end{equation}
\begin{equation}\label{rd_fits2}
Q = 2(1-a)^{-9/20},
\end{equation}
which are accurate to $\approx 5$ per cent. Hughes \& Blandford (2003)
note that the value of $a$ will depend on the role that SMBH merging
plays in SMBH evolution.  We follow Hughes (2002) and adopt $a =
0.997$. The observed strain [$H(t)$] is the sum of the strain in each
polarisation weighted by the corresponding detector response functions
[$F_+(\theta,\phi,\psi),F_{\times}(\theta,\phi,\psi)$; Thorne 1987]
for a source with polarisation axes rotated by an angle $\psi$ to
$\hat{n}(\theta,\phi)$

\begin{equation}
H(t) = F_+(\theta,\phi,\psi)h_+(t) +
F_{\times}(\theta,\phi,\psi)h_{\times}(t).
\end{equation}

For $a\approx1$, the quality factor is large and most of the energy
radiated during the ringdown phase is observed at the central ringdown
frequency ($f_{\rm rd,obs}$). In this limit the SNR may be
approximated by (Hughes 2002)

\begin{equation}\label{ring_SNR}
{\rho}^2=\frac{2 \int_{0}^{\infty} H^2(t)}{S_h(f_{\rm rd,obs})}dt.
\end{equation}

\noindent For comparison with the sky averaged inspiral SNR we average
the SNR in equation (\ref{ring_SNR}) over randomly generated sets of
$(\theta,\phi,\psi,\lambda,\hat{L}.\hat{n})$.

\subsection{Signal to noise ratios}

\begin{figure*}
\vspace*{55mm}
\includegraphics{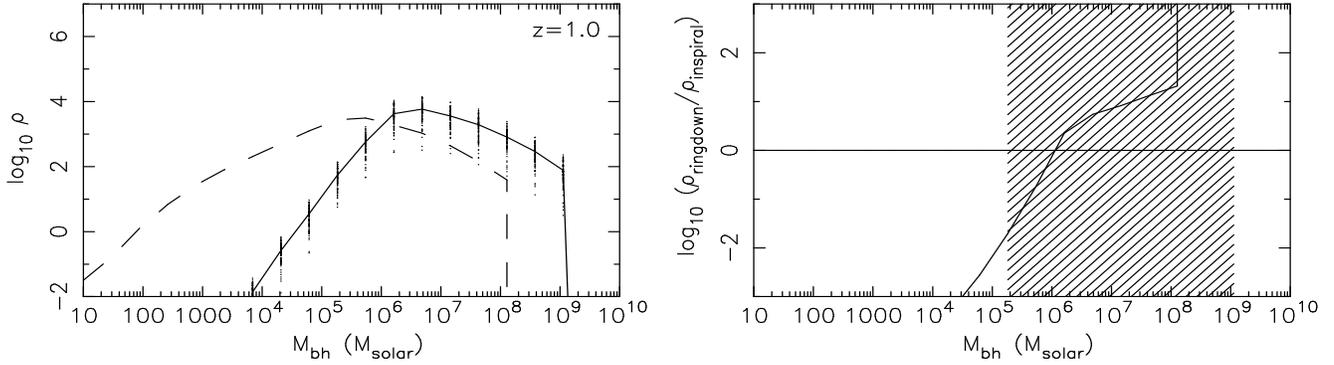}
  \caption{\label{SNR_results} Signal to noise ratios for the inspiral
  (\emph{dot-dashed line}) and ringdown (\emph{solid line}) phases of
  equal mass binary black hole coalescence as a function of single
  black hole mass (left panel). For completeness we show the ringdown
  signal to noise ratios for random sky angles and orientations
  (\emph{dots}). On the right we plot the ratio of ringdown to
  inspiral signal to noise as a function of coalescing black hole mass.
  The hatched region denotes the mass range over which the ringdown
  signal allows confident detection ($\rho \geq 5$). }
\end{figure*}

More massive BBH systems radiate GWs at lower frequencies and for a
given binary system the ringdown signal is at a higher frequency [see
equations (\ref{f_oneyr}) \& (\ref{rd_fits1})].  Therefore if the
sensitivity curve has a low frequency cut off and/or degrades rapidly
with frequency, searches for the ringdown signal will probe more
massive BBHs.  We find that ringdown searches can detect coalescence
between BHs with masses up to $M_{\rm bh} \approx 10^{9.1}$~M$_\odot$
($10^{8.1}$~M$_\odot$) at $z=1$ ($z=9$).  In the left hand panel of
Figure~\ref{SNR_results} we plot the SNR for the inspiral
(\emph{dot-dashed line}) and ringdown (\emph{solid line}) phases of
equal mass, $z=1$ BBH coalescence as a function of single BH mass.
The right hand panel shows the ratio of ringdown to inspiral SNRs,
with the hatched region denoting the mass range over which the
ringdown signal is detectable. The SNR is greater in the ringdown
signal for $M_{\rm bh} > 10^{6}$~M$_\odot$.  We note that there exists
a range of SMBH masses ($M_{\rm bh} \approx 10^{8.2} -
10^{9.1}$~M$_\odot$ at $z = 1$) for which BBH coalescence is
detectable only in the ringdown phase.  Since characteristic
frequencies decay with $(1+z)$, this mass range decreases with
redshift. By $z = 9$, the mass range is comparable to the typical SMBH
mass observed in the local universe ($M_{\rm bh} \approx 10^{7.6} -
10^{8.2}$~M$_\odot$). These results are consistent with the
calculations of FH98 and reinforce the point that searches for
gravitational radiation from the ringdown phase may generate
constraints on the high mass SMBH population.  The higher frequency of
the ringdown signal from SMBH binaries offers limited advantage for
their detection if the low frequency sensitivity has a power-law slope
of $-2$ and no lower cut-off (corresponding to the optimal scenario of
constant accelerometer dominated noise, see White~2002). However, it
is expected that frequency dependent sources of spurious acceleration
will result in a steeper low-frequency sensitivity curve
(Bender~2003).

In addition to the instrumental noise, unresolved GWs from various
classes of binary systems are expected to limit LISA's sensitivity
(see Sesana et al.~2004b).  However, Sesana et al. (2004b) calculate
that BBH coalescence only dominates the sky noise below a frequency of
$10^{-4}$~Hz, and this noise is still an order of magnitude lower than
the suggested LISA sensitivity.  Therefore the unresolved BBH
background should not significantly affect LISA's sensitivity to
supermassive BBH coalescence.

\section{Merger and Coalescence rates}\label{merger}

Estimates of the merger rate of dark matter halos are provided by the
extended Press-Schechter~(1974) formalism (Lacey \& Cole~1993).  We
denote the merger rate per halo of mass $M$ per unit time with halos
of mass between $\Delta M$ and $\Delta M +dM$ by $\frac{d^2N_{\rm
mrg}}{d\Delta Mdt}|_M$. The local (all-sky) detection rate of SMBH
mergers ($\frac{d^2N}{dzdt}$) occurring in the redshift interval
($z,z+dz$) may be estimated (e.g. Wyithe \& Loeb~2003) by integrating
the halo merger rate ($\frac{d^2N_{\rm mrg}}{d\Delta Mdt}|_M$) over
the Press-Schechter dark matter halo mass function ($\frac{dn}{dM}$)
as below:
\begin{eqnarray}
 \label{event_rate}
 \nonumber
\frac{d^2N}{dzdt} &=& \int_{0}^{\infty}dM \int_{M/3}^{M} d\Delta M
 \langle\Theta(M_{\rm bh},\Delta M_{\rm bh},z)\rangle \frac{dn}{dM}\\
&&\hspace{-10mm}\times 
 \left. \frac{d^2N_{\rm mrg}}{d \Delta M dt}\right| _M f_{\rm bh}(M,z) f_{\rm bh}(\Delta M,z) \frac{\epsilon_{\rm mrg}}{1+z}
 4 \pi \frac{d^2V}{dz d \Omega},
\end{eqnarray}

\noindent where $\frac{d^2V}{dz d \Omega}$ is the comoving volume per
 unit solid angle between $z$ and $z+dz$ and the factor $(1+z)^{-1}$
 accounts for time-dilation.  The product of the halo SMBH occupation
 fractions [$f_{\rm bh}(M,z)f_{\rm bh}(\Delta M, z)$] relates the
 expected SMBH merger rate to the halo merger rate.

Equation~(\ref{event_rate}) has the following additional features.
We assume an observationally motivated dependence of SMBH mass on host
halo mass [$M_{\rm bh} = M_{\rm bh}(M,z)$, see
\textsection~\ref{mbh_reln}]. The merger rate $\frac{d^2N}{dzdt}$ is
weighted by the probability for detection of the central SMBH
coalescence, denoted $\langle\Theta\rangle$, where the detection
$\Theta$ is determined through calculation of the SNR
\begin{eqnarray}
\nonumber
\Theta(M_{\rm bh},\Delta M_{\rm bh},z) &=& 0, {\rm if\ } \rho(M_{\rm bh},\Delta M_{\rm bh},z) < 5,\\ 
&=&1, {\rm if\ } \rho(M_{\rm bh},\Delta M_{\rm bh},z) \geq 5.
\end{eqnarray}
\noindent We exclude halo mergers where the accreted satellite halo
takes longer than the Hubble time to sink to the centre of the merged
system, since the BBH will not have time to form. This is achieved by
limiting the merger rate contribution to halo pairs with mass ratios
smaller than 3 (Colpi, Mayer \& Governato~1999).  We assume efficient
coalescence of SMBH binaries in halo mergers, making our prediction of
the coalescence rate a maximum. The merger rate scales with the
fraction of BBHs that evolve into the GW dominated regime
($\epsilon_{\rm mrg}$).  We further assume coalescence to be rapid.
This last assumption may affect the redshift distribution of our
expected counts (although this effect may be small, see Sesana et
al.~2004a), but should have little effect on estimates of the total
event rate.

\subsection{The $\bmath{M_{\rm bh}}$--$\bmath{M_{\rm halo}}$ relationship}\label{mbh_reln}

Shields et al.~(2003) investigated the evolution of the relationship
between $M_{\rm bh}$ and the velocity dispersion of the host bulge
($\sigma_c$) out to $z\sim3$ using a sample of radio-quiet AGN. They
found no evidence for variation from the locally determined
relationship (Ferrarese \& Merritt 2000; Gebhardt 2000)
\begin{equation}
M_{\rm bh} = 10^{8.13} {\rm M}_\odot (\frac{\sigma_c}{200 {\rm
 kms}^{-1}})^{4.02}.
\end{equation}
Ferrarese (2002) found that the velocity dispersion of the local
galactic bulges correlates with the circular velocity in the flat
region of the halos rotation curve $\sigma_c \sim v_c^{1.2}$, indicating
that $M_{\rm bh} \sim v_c^5$. Assuming that there is no evolution
in the $\sigma_c$--$v_c$ relation, and relating $v_c$ to the virial 
velocity of a virialised halo of mass $M$
\begin{eqnarray}
\nonumber 
v_{\rm vir} &=& 23.4 \left( \frac{M}{10^8 h^{-1}{\rm
M}_\odot} \right)^{1/3} \left[ \frac{\Omega_m}{\Omega_m^z}
\frac{\Delta_c}{18\pi^2}\right]^{1/6} \\
&\times& \left( \frac{1+z}{10} \right)^{1/2}{\rm kms}^{-1},
\end{eqnarray}
\noindent
[where
$\Omega_m^z=\frac{\Omega_m(1+z)^3}{(\Omega_m(1+z)^3+\Omega_\Lambda+\Omega_k(1+z)^2}$,
$d \equiv \Omega_m^z-1$ and $\Delta_c=18\pi^2+82d-39d^2$ is the
overdensity of a virialised halo at redshift $z$] we can determine the
redshift dependent $M_{\rm bh}$--$M_{\rm halo}$ relationship
(normalised to the local relationship from Ferrarese~2002)
\begin{equation}\label{mbh-mhalo}
M_{\rm bh} = 10^9 {\rm M}_\odot \left( \frac{M_{\rm halo}}{1.5\times
10^{12} {\rm M}_\odot} \right) ^{5/3} \left[ \frac{1+z}{7}
\right]^{5/2}.
\end{equation}

The $M_{\rm bh}$--$M_{\rm halo}$ relationship has been established
locally for halo masses in the range $10^{11} - 10^{13.4}{\rm
M}_{\odot}$ (Ferrarese~2002).  Fits to gravitational lens separation
distributions [Porciani \& Madau 2000; Kochanek \& White 2001
hereafter KW2001] and simple models for baryon cooling (Cole et
al. 2000; KW2001) indicate that halos with $M \ga 3 \times
10^{12}$~M$_\odot$ tend to form groups and clusters of galaxies.  This
will suppress the number of very massive SMBHs relative to the number
of massive dark matter halos. In addition, limitations on accretion
from the IGM, and feedback from massive galactic winds are expected to
suppress starformation in small mass halos (see Barkana \& Loeb 2001
and references therein), and may also inhibit SMBH formation. In the
following section, we construct an observationally motivated SMBH
occupation fraction ($f_{\rm bh}$) that empirically accounts for these
effects.

\subsection{The black hole occupation fraction}

A model which assumes that all dark-matter halos contain SMBHs (e.g.
Wyithe \& Loeb~2003) may substantially overestimate the SMBH
coalescence rate.  Here we assume that all galaxies residing in halos
with virial temperatures larger than a critical value host a central
SMBH [with mass determined by the $M_{\rm bh}$--$M_{\rm halo}$
relationship, equation~(\ref{mbh-mhalo})].  We construct the galaxy
occupation fraction of dark-matter halos at $z=0$ by comparing the
observed velocity distribution of local galaxies with a velocity
function that is generated from the Press-Schechter mass-function, and
which assumes a relationship between cooled baryonic and virial
velocities (see \textsection \ref{gal_vel}).  Motivated by the absence
of evolution in the observed correlation between $M_{\rm bh}$ and the
velocity dispersion of the host bulge, we assume that the fraction of
halos of mass ($M$) and redshift ($z$) that host a central SMBH is
determined by $v_{\rm vir}$ independently of redshift [ie. $f_{\rm
  bh}(M,z) = f_{\rm bh}(v_{\rm vir})$].

\subsubsection{Galaxy velocity functions and SMBH demography}\label{gal_vel}

Measurements of the circular velocities of galaxies reflect the
dynamics of the stars in the inner region of the halo where the
density profile differs significantly from the profile predicted by
Navarro, Frenk \& White~(NFW, 1997) for CDM.  Simple galaxy formation
models consider the effect of adiabatic cooling of baryons on the
radial density and velocity profiles of galaxies but neglect heating
due to starformation.  Adiabatic cooling and condensation of the
halo's baryonic component into a rotationally supported disk steepens
the inner density profile, resulting in a higher maximal circular
velocity ($v_{\rm cool}$) than implied by the initial profile
(Dalcanton, Spergel \& Summers 1997; Mo, Mao \& White 1998; Gonzalez
et al. 2000; KW2001).

To model the cooled halo density profile, Mo et al.~(1998) assume that
the baryonic and dark matter components are initially uniformly mixed
in an NFW density profile with concentration parameter $c$, spin
parameter $\lambda$ and specific angular momentum $j = 0.05$. The peak
circular velocity of the modified density profile may be determined by
self-consistently solving (with angular momentum and energy
conservation) for the scale factor of the exponential disk.  For a
disk with specific angular momentum equal to that of the original
halo, the cooled velocity is related to the virial velocity by (Mo et
al.~1998)
\begin{eqnarray}\label{vcool}
\nonumber
v_{\rm cool} &\approx& (1-3m_d+5.2m_d^2)\left( \frac{\lambda}{0.1}
\right)^{-0.06+2.71m_d+0.0047/\lambda}\\ 
&\times& (1-0.019c+0.0025c^2+0.52/c)v_{\rm vir}.
\end{eqnarray}
\noindent Numerical N-body simulations of hierarchical structure
formation suggest the average halo spin to be roughly independent of
halo mass and equal to about $0.04$ (Bullock et al.  2001a).
Furthermore, the concentration parameter of the NFW profile may be
related to the halo virial mass $M_{\rm vir}$ (at $z=0$) by $c= 9
\left( \frac{M_{\rm vir}}{8.12 \times 10^{12} {\rm M}_{\odot}} \right)
^{-0.14}$ (Bullock et al.  2001b).  We assume the fraction of halo
mass in the cooled disk to be $m_d = 0.05$, independent of halo
mass. The cooled baryon fraction is thought to be smaller in more
massive halos (see KW2001), however we find that allowing $m_d$ to
vary over halo mass as suggested by the cooling model of KW2001 has
little effect on the profile of the adiabatically cooled velocity
function.  The halo cooled velocity function may then be computed from
the PSMF through the change of variable
\begin{equation}\label{gal_func}
\frac{dn}{dv_{\rm cool}} = \frac{dn}{dM} \left| \frac{dM}{dv_{\rm
cool}} \right|.
\end{equation}
\noindent For a cooled baryon fraction ($m_d$) that is independent of
mass, this prescription results in a $M_{\rm halo} \sim v_{\rm
cool}^3$ dependence of cooled velocity on halo mass for $10^6$~M$_\odot
< M_{\rm halo} < 10^{15}$~M$_\odot$. Consequently, the net effect of
this adiabatic cooling model is to translate (and re-scale) the
Press-Schechter distribution of halo virial velocities.  To construct
our occupation fraction we assume that there is one circular velocity
for which galaxies have an occupation fraction of unity. We then find
the (linear) relationship between cooled and virial velocity that
enforces this condition.  We have adopted the observational galaxy
velocity function used by KW2001 for this purpose.  KW2001 provide
data down to $v_{\rm cool} \approx 70$~km~s$^{-1}$. Below this we
extrapolate with a single power law $\frac{dn}{dv_{\rm cool}} \sim
v^{\beta}$ with $\beta = -1$ (appropriate for a luminosity function
with faint-end slope $-1$).

\begin{figure}
\vspace*{77mm}
\includegraphics{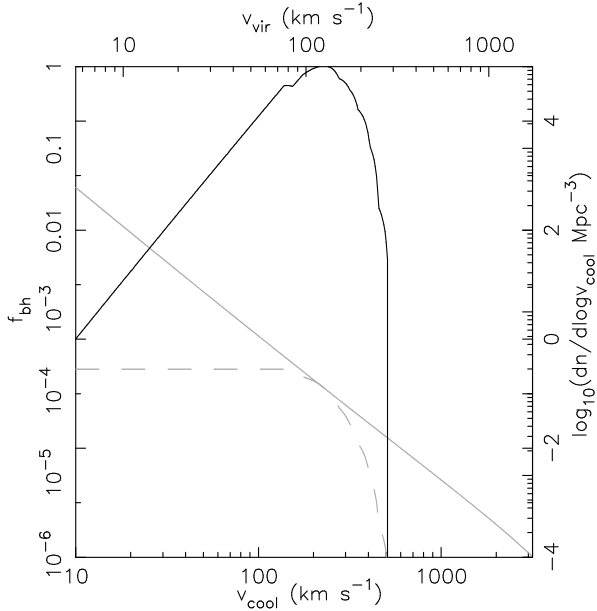}
\caption{\label{fig2} Occupation fraction (\emph{solid black line}) of
  SMBHs in halos as a function of the maximum circular velocity of the
  cooled density profile (lower axis) or the halo virial velocity
  (upper axis).  The \emph{solid grey} and \emph{dashed grey} lines
  show the (local) cooled Press-Schechter halo velocity distribution
  and observed galaxy velocity distribution (calibrated by the right
  hand axis) as described in \textsection\ref{gal_vel}. }
\end{figure}

This process returns a halo cooled velocity ($v_{\rm cool}$) which, at
the peak of the occupation fraction, is consistent with the velocity
derived using the prescription of Mo et al.~(1998) to within $\approx
15$ per cent.  Moreover, our relationship is consistent with the ratio
of $v_{\rm vir}/v_{\rm cool} \approx 1.8$ measured from weak lensing
studies (Seljak~2002).  The resulting occupation fraction (see
Figure~\ref{fig2}) peaks at $v_{\rm cool} \approx 230$~km~s$^{-1}$
($v_{\rm vir} \approx 125$~km~s$^{-1}$).  This cooled velocity is also
consistent with the circular velocity at which Press-Schechter and
observed galaxy velocity functions agree in the works by Gonzalez et
al.~(2000) and KW2001.

van den Bosch et al.~(2002) provide another estimate of the dependence
of galaxy formation efficiency on host halo mass via the average halo
mass-to-light ratio.  This quantity may be estimated by constructing
halo mass conditional luminosity functions using the relationship
between (luminosity dependent) observed galaxy-galaxy correlation
lengths and mass-scale for a hierarchical cosmology (see van den Bosch
et al.~2002).  Their calculation yields a minimum mass-to-light ratio
(indicative of maximum galaxy formation efficiency) for dark matter
halos with $M_{\rm halo} \sim 3 \times 10^{11}$~M$_\odot$, which is
comparable to the peak in $f_{\rm bh}(M,z=0)$ at $M \sim 4.8 \times
10^{11}h^{-1}$~M$_\odot$.

\begin{figure}
\vspace*{93mm}
\includegraphics{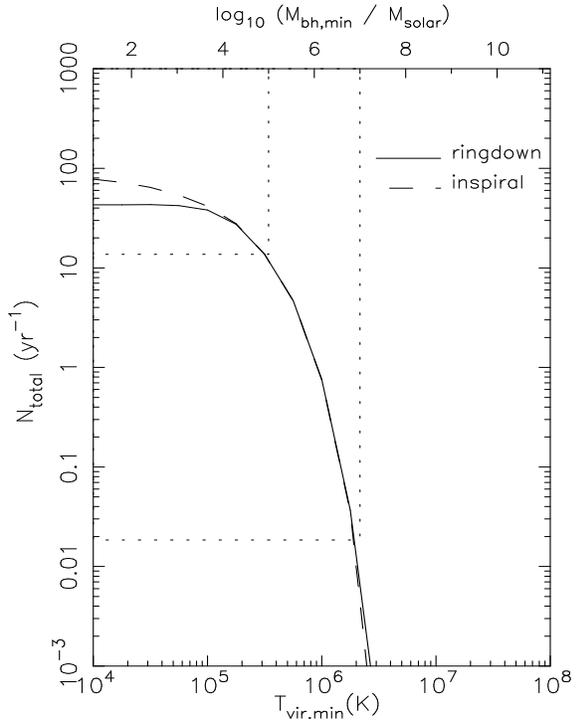}
\caption{\label{fig3} Expected detectable binary black hole
coalescence rates as a function of the minimum virial temperature of a
black hole hosting halo ($T_{\rm vir,min}$, lower axis) or the minimum
black hole mass ($M_{\rm bh,min}$, upper axis) for both inspiral
(\emph{dashed line}) and ringdown (\emph{solid line}) phases assuming
efficient coalescence. The \emph{dotted lines} highlight the predicted
event rates if the minimum SMBH mass is taken to be the minimum SMBH
mass in local galactic bulges ($M_{\rm bh,min} \approx
10^{5}$~M$_\odot$) or the typical mass of a SMBH powering a quasar at
$z \sim 2$ ($M_{\rm bh,min} \approx 10^{7}$~M$_{\odot}$).}
\end{figure}
The occupation fraction decays rapidly for cooled velocities larger
than $230$~km~s$^{-1}$.  We note that errors in $f_{\rm bh}$ at high
$v_{\rm vir}$ will not effect our predicted event rates due to the
exponential tail of the PSMF. Indeed Wyithe \& Loeb~(2003) found that
imposing a sharp cut-off to $f_{\rm bh}$ at a virial velocity of
$600$~km~s$^{-1}$ had little impact on their predictions.

\begin{figure*}
\vspace*{65mm}
\includegraphics{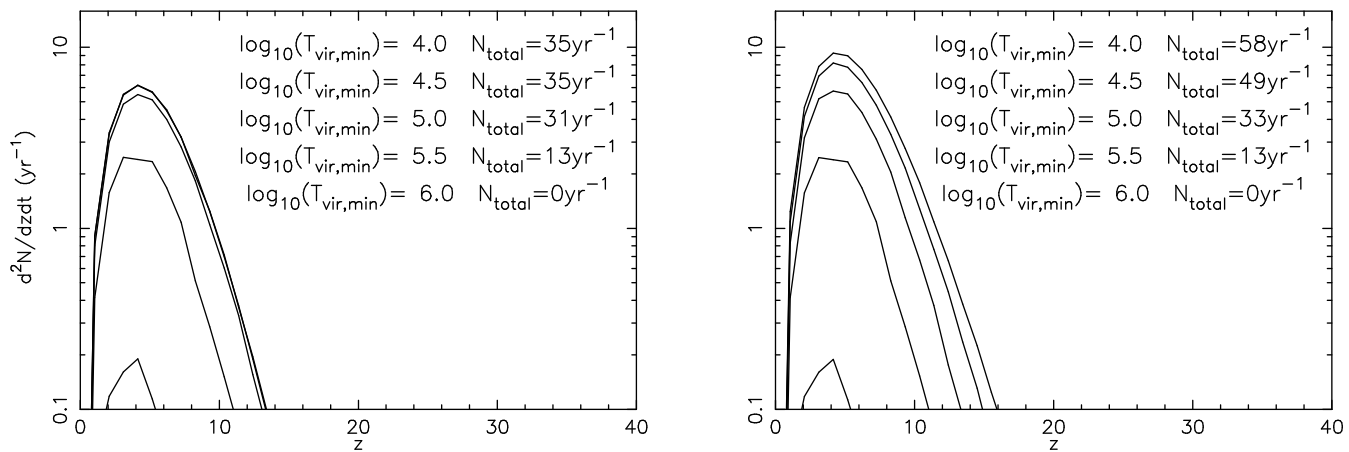}
\caption{\label{fig4}Redshift distribution of ringdown (left panel)
and inspiral (right panel) phase detections of binary black hole
coalescence for several $T_{\rm vir,min}$ assuming rapid, efficient
coalescence. The total event rates ($N_{\rm total}$) for each $T_{\rm
vir,min}$ curve are listed in order of descending peak detection
rate.}
\end{figure*}

\subsubsection{The minimum SMBH mass}

Prior to the reionisation of the IGM, halos required a virial
temperature larger than $\sim10^4$~K for the accreted gas to cool via
atomic hydrogen transitions within a Hubble time.  After reionisation
of the IGM, the higher IGM temperature inhibited the accretion of gas
into halos with virial temperatures lower than $\sim2\times10^5$~K.
These virial temperatures specify the minimum mass of a halo inside of
which a SMBH may form (e.g. Barkana \& Loeb~2001)
\begin{eqnarray}\label{Mvir}
\nonumber
M_{\rm vir} = 10^8 h^{-1} {\rm M}_\odot \left( \frac{T_{\rm
vir}}{1.98 \times 10^4} \right)^{3/2}\left[ \frac{\mu}{0.6} \right]
^{-1.5}\\ \times \left[
\frac{\Omega_m}{\Omega_m^z}\frac{\Delta_c}{18\pi^2}\right]
^{-1/2}\left( \frac{1+z}{10} \right)^{-3/2},
\end{eqnarray}
\noindent Equivalently, a minimum virial temperature specifies a
minimum depth for the potential well (or velocity, or mass) of a halo
in which a SMBH can exist.  Given the $M_{\rm bh}$--$M_{\rm halo}$
relationship, equation~(\ref{Mvir}) implies a minimum SMBH mass
($M_{\rm bh,min}$) which is independent of redshift.  For $T_{\rm
  vir,min} = 10^4$~K~($2\times10^5$~K) this minimum SMBH mass
corresponds to $M_{\rm bh,min} \approx 15$~M$_\odot$~($1.5\times
10^4$~M$_\odot$).  In contrast, the locally detected population of
SMBHs reside in more massive halos with virial temperatures greater
than $\approx 10^{5.4}$~K, although this lower limit may be partially
attributable to the difficulties in obtaining dynamical evidence for
BHs with $ M_{\rm bh} \lesssim 10^6~{\rm M}_{\odot}$ (e.g.
Kormendy \& Richstone~1995).

\subsection{Event-rate predictions}

The predicted event rates are plotted in Figure~\ref{fig3} as a
function of the minimum virial temperature (lower axis) and minimum
SMBH mass (upper axis).  The event rates per year decline with
increasing $T_{\rm vir,min}$. For $T_{\rm vir,min} \approx 10^5$~K
($M_{\rm bh,min}\sim10^4$~M$_\odot)$ we predict $\approx 40$
detections per year in each of the ringdown and inspiral phases. These
predictions are significantly below those of Wyithe \& Loeb~(2003),
with the discrepancy due to our inclusion of the SMBH occupation
fraction in the merger rate.  Our SMBH occupation fraction and
accurate detection criteria result in predictions that decay more
rapidly with increasing $T_{\rm vir,min}$ than previous calculations
suggested.

SMBHs powering luminous quasars at $z \sim 2$ weigh
$\ga10^7$~M$_\odot$. If $M_{\rm bh,min} \sim 10^{7.1}$~M$_{\odot}$ the
predicted event-rates are very low; falling below $0.001$~yr$^{-1}$
for each phase.  Hence LISA will not detect coalescence from the major
merges of bright quasar hosts.  However, there is evidence for SMBHs
as small as $\approx 10^5$~M$_\odot$ in local galactic bulges
(Ferrarese~2002) and low-luminosity, low-redshift AGN (Greene \&
Ho~2004). A reasonable estimate for the minimum SMBH mass is therefore
$M_{\rm bh,min} = 10^5$~M$_\odot$, which corresponds to predictions of
$\approx 15$ events per year in each phase.  Figure~\ref{fig4} shows
the redshift distribution of events as a function of the minimum
virial temperature. Most of the events originate at $z \approx 3 - 4$
due to mergers between halos more massive than $10^{10}$~M$_\odot$.

Sesana et al.~(2004a\&b) make predictions for the detection rate of
inspiral phase BBH coalescence for a seeded SMBH formation model;
estimating $\approx 35$ ``bursts'' in 3 years of observation due to
events at $z \approx 4$.  This calculation employs a specific model
for the formation and evolution of BBHs embedded in an isothermal
stellar distribution. Sesana et al.~(2004a) use the seeded SMBH growth
model of Volonteri et al.~(2003) in which (large) stellar mass BHs
form in $3.5-\sigma$ peaks at $z=20$ and grow via accretion and
coalescence during major halo mergers.

While we also rely on the merger rate of dark matter halos for the
basis of our coalescence rate, our calculation differs by employing
empirically motivated estimates for the population of SMBHs in
halos. We do not assume the physics leading to the absence of
galaxies (and by extension SMBHs) in large and small halos, however we
have shown that the inclusion of the occupation fraction significantly
affects the coalescence rate. We have assumed an efficient coalescence of SMBH
binaries, while Sesana et al.~(2004a) specify a detailed model for
binary evolution that results in some binaries being ejected from the
galaxy core (where they do not coalesce). 

We estimate an upper limit for the rate of supermassive BBH
coalescence (where both binary members weigh more than
$10^5$~M$_\odot$) of $\sim 15$~yr$^{-1}$, which is larger than, but
comparable to, the $\sim 9$~yr$^{-1}$ estimated by Sesana et
al.~2004a\&b for binary coalescence with one member heavier than
$10^5$~M$_\odot$. If intermediate mass black holes exist, then our
estimate for the detectable coalescence rate increases only modestly
[due to the declining occupation fraction for lower mass halos (see
Figure~\ref{fig2})] to $\sim$ tens per year (see Figure~\ref{fig3}).

\section{Conclusion}
\label{concl}

We have computed the SNR of both the inspiral and ringdown phases of
SMBH coalescence, and calculated mass and redshift ranges over which
BBH coalescence will be detectable by LISA.  The higher frequency of
the ringdown signal allows detection of the coalescence of more
massive binary systems.  In particular, searches for gravitational
waves for the ringdown signal offer a superior probe of BBH
coalescence between SMBHs which are more massive than $\approx
10^6$~M$_\odot$, and allow confident detection of the coalescence of
SMBHs in the mass range $M_{\rm bh} \approx 10^{8.2} -
10^{9.1}$~M$_\odot$ at $z=1$ should they exist (or $M_{\rm bh} \approx
10^{7.6} - 10^{8.2}$~M$_\odot$ at $z=9$).
 
We have predicted the rate of binary black hole coalescence detectable
by LISA assuming a hierarchical cosmology in which black holes form in
halos with virial temperature above a critical value and obey an
observationally motivated scaling with halo mass and redshift of the
form $M_{\rm bh} \sim M_{\rm halo}^{5/3}(1+z)^{5/2}$. Prompted by the
observation that the numbers of both large and small galaxies fall
short of the expected numbers of dark-matter halos, we compute the
empirical galaxy occupation fraction. This is obtained by comparing
the locally observed galaxy velocity function with a cooled
Press-Schechter~(1974) velocity function. Our predicted event rate
declines rapidly with increased estimates for the smallest SMBH that
can form.  If we are guided by the smallest nuclear BH mass inferred
in local galactic bulges and nearby low-luminosity AGN ($\approx
10^5$~M$_\odot$), then we expect $\approx 15$ detections per year.
Most of these events will be detectable via both their final-year
inspiral and ringdown signals. Our observationally motivated
estimates for the event rate of BBH coalescence therefore indicate
that observations of coalescing BBHs are a realistic target for LISA.

\section*{Acknowledgements}

The authors wish to thank Rachel Webster, Andrew Melatos and Avi
Loeb for helpful discussions during the course of this work.

\bsp

\label{lastpage}

\end{document}